# Ferromagnetism and correlated insulating states in monolayer $Mo_{33}Te_{56}$

Zemin Pan[1#], Wenqi Xiong[1#], Jiaqi Dai[2,3#], Hui Zhang[1#], Yunhua Wang[4], Tao Jian[1], Xingxia Cui[1], Jinghao Deng[1], Xiaoyu Lin[1], Zhengbo Cheng[1], Yusong Bai[1], Chao Zhu[1], Da Huo[1], Geng Li[5,6], Min Feng[1], Jun He[1,7], Wei Ji[2,3]*, Shengjun Yuan[1,7]*, Fengcheng Wu[1,7]*, Chendong Zhang[1,7]*, and Hong-Jun Gao[5]

[1]*School of Physics and Technology, Wuhan University, Wuhan 430072, China*

[2]*Beijing Key Laboratory of Optoelectronic Functional Materials & Micro-nano Devices, Department of Physics, Renmin University of China, Beijing 100872, China*

[3]*Key Laboratory of Quantum State Construction and Manipulation (Ministry of Education), Renmin University of China, Beijing, 100872, China*

[4]*Key Laboratory of Quantum Theory and Applications of Ministry of Education and School of Physical Science and Technology, Lanzhou University, Lanzhou 730000, China*

[5]*Beijing National Center for Condensed Matter Physics and Institute of Physics, Chinese Academy of Sciences, Beijing 100190, China*

[6]*Hefei National Laboratory, Hefei 230088, China*

[7]*Wuhan Institute of Quantum Technology, Wuhan 430206, China*

**Correspondence and request for materials should be addressed to:*
wji@ruc.edu.cn (W.J.), s.yuan@whu.edu.cn (S.J.Y), wufcheng@whu.edu.cn (F.C.W), cdzhang@whu.edu.cn (C.D.Z)

**Abstract: Kagome lattices have an inherent two-dimensional nature. Despite previous realizations in the monolayer limit, their abilities to drive emergent electronic states such as correlated insulators have remained unobserved. Here, we report the experimental realization of a new structural phase of monolayer $Mo_{33}Te_{56}$, characterized by its virtually global uniformity as a mirror-twin boundary loop superlattice embedded in an H-$MoTe_2$ monolayer. Through a combination of scanning tunnelling microscopy (STM) and theoretical calculations, we unveil a kagome geometry along with multiple associated sets of kagome flat bands. Crucially, the partial filling of these kagome bands induces ferromagnetism as revealed by spin-polarized STM, and leads to a correlated insulating state exhibiting a hard gap as large as 15 meV. Our findings represent a major advance in kagome materials, offering a framework with clearer band structures and more intrinsic two-dimensional properties for exploring flat-band physics.**

Suppressing electron kinetic energy in solids by reducing band dispersion, ultimately forming flat bands (FBs), is a promising route to enhance many-body correlation effects[1]. Unlike the mundane cases attributed to atomic orbital localizations, wavefunction overlapping and electron hopping processes are preserved in kagome flat bands[2], where the suppression of kinetic energy arises from destructive interference effect[2,3]. Experimentally, several transition-metal compounds, such as $Fe_3Sn_2$, $Co_3Sn_2S_2$, and $YMn_6Sn_6$, are recognized as kagome magnets, given that transition metal atoms form corner-shared triangular sublattices[4-9]. However, the non-negligible hopping along the direction out of the kagome plane complicates the electronic bands in these bulk materials, obscuring the manifestation of the distinct characteristics of FBs[8-10].

Given the two-dimensional (2D) nature of kagome physics and the enhanced *e-e* interactions resulting from reduced dimensionality, atomically thin van der Waals (vdW) materials that mitigate interlayer couplings provide an ideal platform for nearly pristine manifestations of flat-band physics. In artificially constructed multilayers, such as twisted graphene and transition metal dichalcogenides, flat bands are obtained by creating nanometer-size moiré supercells that fold and flatten the original band structure[11-15]. These flat bands, residing across or being tuned to the Fermi level ($E_F$), give rise to correlated insulating states[15-17], magnetism[17,18], and strong-coupling superconductivity[19,20]. However, these flat-band-driven correlated phenomena are highly dependent on the precision of the stacking process, posing significant challenges for potential technological applications. Moreover, the large unit cells and low electron densities intrinsic to these twisted moiré layers tend to result in low-energy, delicately interacting systems. This highlights the pressing need to realize flat bands near $E_F$ in non-moiré 2D crystals[2].

Despite extensive theoretical[2,21] and experimental efforts[22,23], exotic correlated phenomena driven by 2D kagome lattices remain elusive in stoichiometric single-layer compounds. In this work, we report a $Mo_{33}Te_{56}$ phase derived from monolayer H-$MoTe_2$, featuring a thermodynamically stable mirror twin boundary (MTB) loop superlattice with a uniform lattice constant of 19.1 Å. This uniformly replicated structure virtually across the sample highlights a previously unreported stoichiometric phase. Combining scanning tunnelling microscopy (STM), noncontact atomic force microscopy (nc-AFM), and theoretical calculations, we reveal that the formation of an MTB loop supercell gives rise to an electronic kagome lattice with two sets of kagome bands mixed across the $E_F$. The partially filled kagome bands manifest remarkable correlated features, notably itinerant ferromagnetism accompanied by a correlated hard gap of approximately 15 meV in width at the $E_F$. Through advanced applications of spin-polarized STM, we discerned a distinct ferromagnetic hysteresis loop with an effective field measured at 0.1 T (0.35 K). The spatial distribution of spin density and Stoner criteria calculations support that the spontaneous spin-polarization is primarily driven by the high density of states at the $E_F$, contributed by the partially filled kagome flat bands. This work demonstrates a route for materializing exotic FB-driven quantum phases in a vdW monolayer via straightforward and scalable structural engineering.

Using molecular beam epitaxy (MBE) technique, we first synthesized H-phase monolayer $MoTe_2$ on bilayer graphene (BLG)/SiC (0001) [see Methods]. Following synthesis, the sample underwent annealing at 450 °C, which induced the formation of a dominantly ordered superstructure phase within the monolayer (Fig. 1a). Magnified images of this superstructure (Fig. 1b, 1c) show a wagon-wheel-like morphology, characterized by six lines (spokes of the wagon-wheel) encircled in a skewed hexagonal pattern. This superlattice exhibits a periodicity of 19.1 Å (red dashed lattices in Fig. 1b and 1c), with the spoke lines dividing the atomic layer into two types of triangular domains (as elucidated later), marked by green- and orange-dashed triangles. At a high sample bias (2.0 V; Fig. 1b), the line boundaries appear as bright stripes, while at a low sample bias (0.8 V; Fig. 1c), they manifest as dark trenches. These morphologic characteristics are consistent with the STM observations of MTBs-arranged wagon-wheel patterns in H-phase TMDs[24-28]. The possibility of this pattern being a moiré superlattice[29], was also excluded, as it shows no dependence on the twist angle relative to the BLG substrate (Fig. S1).

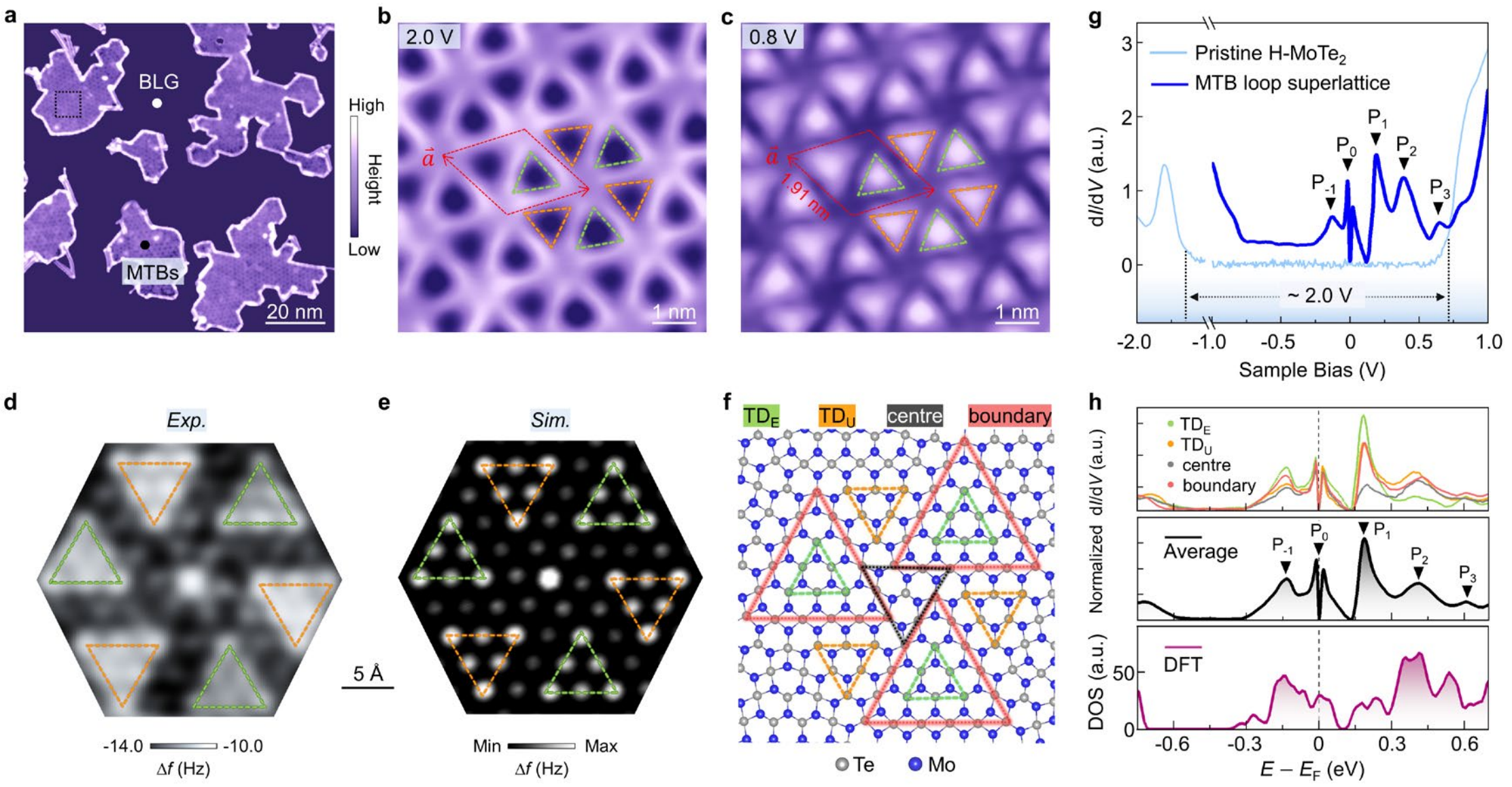

**Fig. 1 | The atomic structure and tunnelling differential conductance spectrum of the ML-$Mo_{33}Te_{56}$.** **a**, Large-scale STM topographic image of the sample hosting the dominant portion of the MTB loop superlattice. **b**, **c**, Zoomed-in STM images of the marked region in **a**. The supercell is labelled a red rhombus with a lattice constant of 1.91 nm. **d**, Atomically resolved nc-AFM images acquired in repulsive

regime (details in Fig. S2). Only the top-layer Te atoms are shown. **e**, The DFT simulation of the frequency shift image [Method]. **f**, Atomic model of the MTB loop superlattice. Grey and blue balls represent Te and Mo atoms, respectively. The boundary-formed triangular loops, $TD_E$, $TD_U$, and centre sites are marked as shown. **g**, Experimental d*I*/d*V* spectra acquired on pristine H-$MoTe_2$ (light-blue) and a typical spectrum of the $Mo_{33}Te_{56}$ (blue). Five pronounced peaks are labelled $P_{-1}$, $P_0$, $P_1$, $P_2$, and $P_3$. **h**, Upper panel: d*I*/d*V* spectra taken at four high-symmetry sites ($TD_{E/U}$, centre, and boundary) and their average spectra are shown in the middle panel. The calculated DOS with spin-polarization and SOC (corresponding to Fig. S6b) is shown in the bottom panel. Scanning parameters: (**a**) bias voltage $V_{bias}$ = 1.0 V, tunnelling current $I_t$ = 10 pA, (**b**) 2.0 V, 50 pA, (**c**) 0.8 V, 50 pA. The stabilization parameters in **g** are $V_{bias}$ = 1.0 V, $I_t$ = 100 pA, and lock-in modulation $V_{mod}$ = 8 mV; in **h** are $V_{bias}$ = 0.7 V, $I_t$ = 200 pA, $V_{mod}$ = 6 mV.

Frequency-shift nc-AFM image (Fig. 1d, acquired in the repulsive regime), primarily capturing surface Te atoms[25,30], provides further information on the atomic structures. The triangular domains (green and orange) match those in the STM images (Fig. 1b and 1c), showing the same wagon-wheel morphology[24-28]. The triangular domains (orange and green) match those in the STM images (Fig. 1b and 1c), showing the same wagon-wheel morphology. There are discernible contrasts between the Te atoms at the boundary and those in the domain regions, which can be attributed to differences in their relative heights and/or local electron densities[25,30]. Each triangular domain consists of six surface Te atoms (*i.e.*, three Mo-Te units) with an interatomic distance of 3.65 Å, which is nearly identical to that in pristine H-$MoTe_2$[27], and there is one Te atom at the rotational centre. Based on these observations, we propose an atomic structure for the observed superlattice, as schematically shown in Fig. 1f[31]. The simulated frequency-shift image (Fig. 1e, [Methods]) based on this model shows good agreement with the experimental morphology. The validity of this model is further verified by its lowest formation energy among various candidate models in a specific range of the Te chemical potential (Fig. S3). In this structure, Te atoms on the MTB form closed triangular loops (marked by red dashed triangles in Fig. 1f), previously referred to as MTB loops[32]. By comparing experimental and theoretical bias-dependent d*I*/d*V* maps (Fig. S4 and Fig. 2), we identify the green triangle domains in the STM and AFM images as enclosed by individual MTB loops, denoted as $TD_E$. The unenclosed orange triangular domains are referred to as $TD_U$. The corner joint region of three MTB loops is denoted as the centre site, highlighted by a black-dashed triangle.

Uniformity was not observed in previous MTB-arranged wagon-wheel structures, leading to the belief that they were line defects. In contrast, the present pattern exhibits uniformity across nearly the entire monolayer sample[24-28]. This uniformity leads to the formation of a new stoichiometric Mo-Te compound, akin to the recently discovered $Mo_5Te_8$ phase[22,33]. Based on our proposed atomic model, the stoichiometry is determined to be $Mo_{33}Te_{56}$ (*i.e.*, $MoTe_{2-x}$ with $x = 10/33$). Figure 1g plots the differential conductance spectra (d*I*/d*V*) of both $Mo_{33}Te_{56}$ and H-$MoTe_2$ monolayers for comparison. Differing from the large band gap observed in H-$MoTe_2$[27], the $Mo_{33}Te_{56}$ spectrum exhibit at least five pronounced peaks, labelled $P_{-1}$, $P_0$, $P_1$, $P_2$, and $P_3$ in ascending order of energy. Notably, peak $P_0$ is centered around the Fermi level along with a zero-bias dip feature. The spatially dependent d*I*/d*V* spectra of four representative sites (upper panel of Fig. 1h) indicate that these primary spectroscopic features are global properties of the system, rather than being attributed to local defects as observed in previous studies on MTBs[24,28]. This distinction also sets our system apart from the honeycomb domain-wall network proposed in the 1T-$TaS_2$[34] (Supplementary Note 1). In Fig. S5, we thoroughly examine the transition of local boundary states into delocalized states as the interboundary separation decreases.

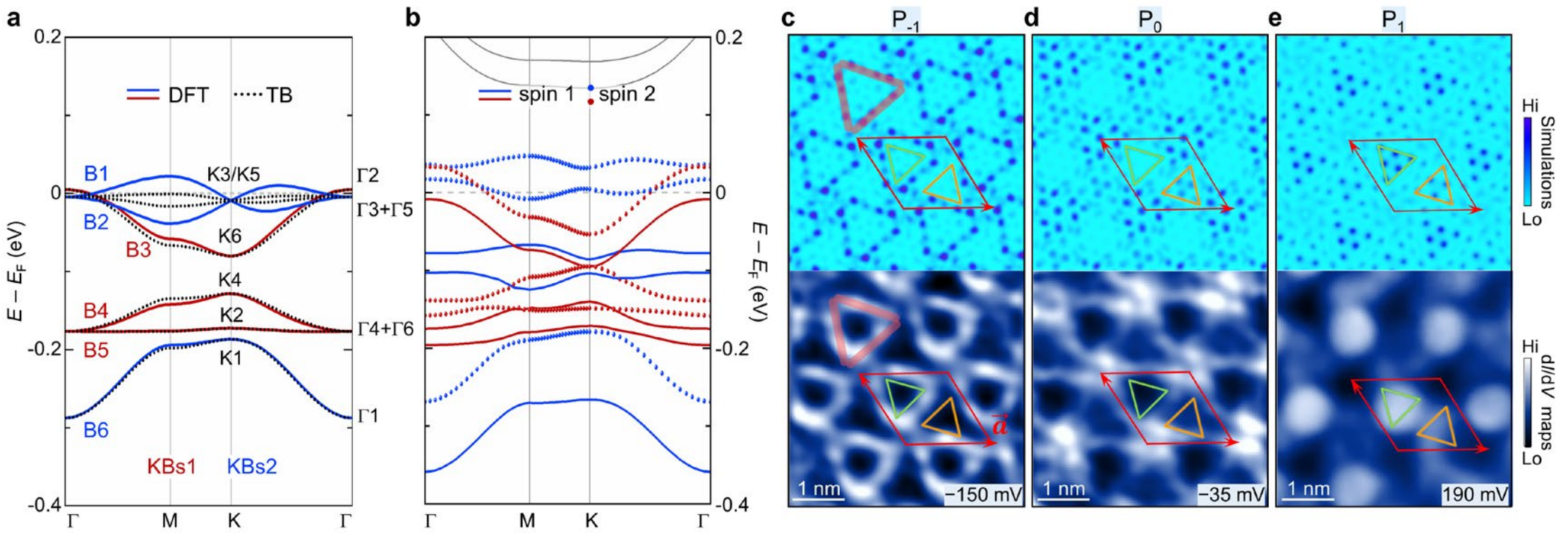

**Fig. 2 | Characterizations of the kagome geometry: DFT analysis, tight binding simulation, and experimental mapping. a**, Calculated band structures of the MTB supercell are plotted along high-symmetry momentum paths in the Brillouin zone near the Fermi level. The wine-red and blue solid lines mark states that are odd (KBs1) and even (KBs2), respectively, under mirror operation $\widehat{M}$. The irreps of states at Γ and K momenta are labelled for each band in **a** (Supplementary Table S1). **b** is the corresponding results with the spin polarization and SOC. The solid and dashed lines mark two spin components that are spin-1 and spin-2, respectively. **c**-**e**, Simulated (upper panel) and experimental (lower panel) d$I$/d$V$ maps for $P_{-1}$, $P_0$ and $P_1$. The energy ranges utilized in these simulations are set at ± 5 mV. More details refer to Fig. S4. The boundary loop is depicted by the pink shadows in **c**. The green and orange triangles in **c**-**e** represent the $TD_E$ and $TD_U$ regions, respectively. The red rhombus in **c**-**e** indicates a supercell. The scanning parameters: **c,** $V_{bias}$= -150 mV, $I_t$ = 200 pA, $V_{mod}$ = 2 mV; **d**, $V_{bias}$ = -35 mV, $I_t$ = 200 pA, $V_{mod}$ = 1 mV; **e**, $V_{bias}$ = +190 mV $I_t$ = 100 pA, $V_{mod}$ = 4 mV.

Our density-functional theory (DFT) calculations reveal the spin-non-polarized and spin-polarized (with spin-orbit coupling, SOC) band structures of the $Mo_{33}Te_{56}$ monolayer (Fig. 2a, 2b and Fig. S6b), in which an on-site Coulomb energy $U$ = 0.5 eV (see detailed discussion on the choice of the $U$ value in Fig. S6) was considered. To establish the resemblance between our DFT results and STS measurements, we plot the density of states (DOS) profile extracted from the spin-polarized (with SOC) bandstructures (see Fig. S4) in the bottom panel of Fig. 1h, which yields a reasonably good agreement with the averaged experimental d$I$/d$V$ curve (middle panel of Fig. 1h). Specifically, the theoretical DOS reproduces the spectroscopic peaks $P_{-1}$ to $P_3$ and captures the dip/gap features observed between $P_1$ and $P_0$, and below $P_{-1}$. The spatial distributions of $P_{-1}$ – $P_3$ states also show a notable consistency between our DFT results and experimental measurements (Fig. 2c-2e and Fig. S4).

These sharp maxima in the DOS profile correspond to a cascade of narrow-width bands, indicative of FBs. To elucidate their physical origins, we adopt a simplified approach by examining a spin-non-polarized band structure focused around $P_{-1}$ and $P_0$ (Fig. 2a), which provides an illustrative understanding of the overall physical scenario. In the spin-degenerate case, six bands near the Fermi level with a relatively simple structure are labelled B1–B6 in descending order of energy. Based on the symmetry analysis, we identify the origin of these narrow-width bands as two sets of kagome bands, as follows. The MTB loop lattice has the $P\bar{6}$ (No. 174) space group and the $C_{3h}$ point group, which includes a threefold rotation $\hat{C}_{3z}$ around the out-of-plane $\hat{z}$ axis and a mirror symmetry $\widehat{M}$ by the $xy$ plane. Bands B3, B4, and B5, referred to as kagome band set 1 (KBs1, in wine red), are odd under the $\widehat{M}$ operation (see details in Supplementary Table S1), while bands B1, B2, and B6, referred to as kagome band set 2 (KBs2, in blue), are even. Bands B3, B4, and B5 (KBs1) exhibit prototypical energy dispersions of a kagome lattice without inversion symmetry, forming a breathing kagome lattice. Band B5 is nearly flat (with a width of 0.4 meV) and touches the B4 band at the Γ point (Brillouin zone centre), and bands B3 and B4 are separated by an energy gap at the K point (Brillouin zone corner).

The formation of an effective breathing kagome lattice by these three bands can be theoretically established using the band representation theory[35,36]. The calculated irreducible representations (irreps) of these three bands are $\Gamma_2 + \Gamma_4 + \Gamma_6$ and $K_2 +$ $K_4 + K_6$ at the Γ and K points, respectively. These symmetry analysis results match well with those for bands formed by Wannier orbitals placed at the $3j$ Wyckoff positions, which correspond to three inequivalent sites per unit cell related by threefold rotation[37]. Geometrically, the $3j$ Wyckoff positions indeed form an effective kagome lattice. In the real space, the effective breathing kagome lattice is further verified. The $P_{-1}$ state, whether spin-degenerate or spin-polarized, is exclusively contributed by KBs1, with its wave-function norm distributed around the MTBs (Fig. 2c). The centres of the MTBs naturally form a kagome lattice akin to the line graph of a honeycomb lattice[21]. Using this schema, our tight-binding (TB) calculations, with fitted hopping parameters and on-site energies, well reproduce the electronic band structures obtained from our DFT calculations (Fig. 2a). Detailed methodology of the TB simulations is documented in Fig. S7 and Supplementary Note 2.

The spin-degenerate band structures (Fig. 2a) reveal that the Fermi level intersects bands B1, B2 and B3, leading to partial fillings of these bands with a total filling factor of 2/3 in the free-standing monolayer. These bands host 4 electrons with bands B1 and B2 occupied by approximately 2.5 electrons and band B3 by approximately 1.5 electrons per unit cell. The partial filling results in a high density of states at the $E_F$ $[D(E_F) = 5.7]$ destabilizing the Fermi surface. The Stoner criterion (ST), calculated as $ST = I \cdot \rho(E_F)$ with the Stoner parameter $I$=0.49, was determined a value of 2.79, suggesting a spontaneously itinerant ferromagnetic (FM) state in monolayer $Mo_{33}Te_{56}$. Although the $I$ value is moderate compared to those of other itinerant magnets like $Fe_3GeTe_2$ (FGT) where $I_{FGT}$ = 0.72[38], the significantly larger DOS compared to FGT (1.5) and FeTe (4.9)[39] is the primary reason for the large ST value. This suggested FM state is corroborated by our DFT results, showing that the FM state is more stable than the non-magnetic state by 11.2 meV/supercell (with SOC) or 24.5 meV/supercell (without SOC). The spin-exchange interaction and SOC cause a spin splitting of approximately of 0.1 eV, as observed in B6 at the Γ point (Fig. 2b), and detailed in Fig. S8.

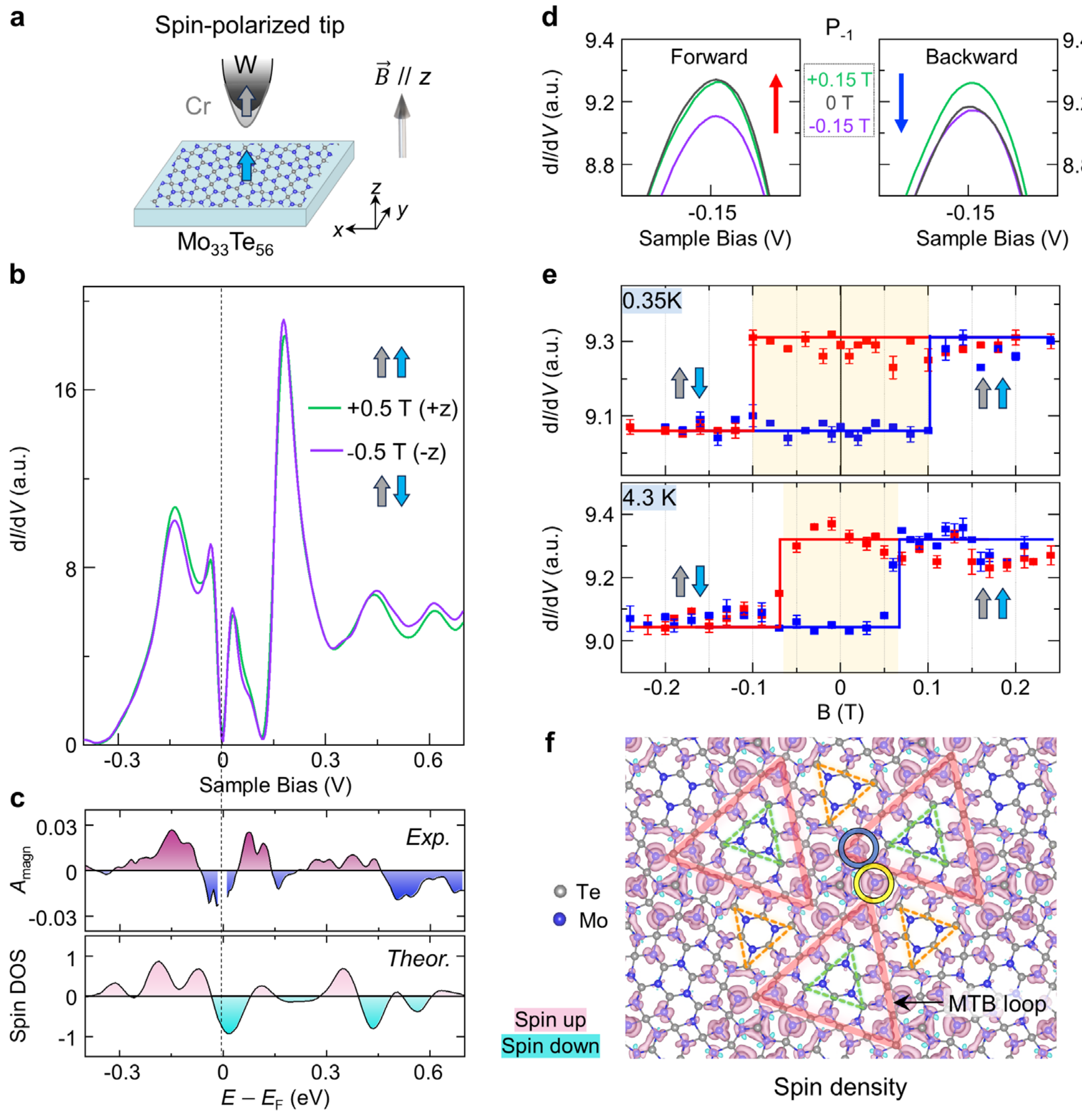

**Fig. 3 | Ferromagnetism in the monolayer $Mo_{33}Te_{56}$. a**, Schematic of SP-STM measurement on ML-$Mo_{33}Te_{56}$ using a Cr-coated W tip with a perpendicular magnetic field. The grey and blue arrows represent the magnetization directions of the tip and sample, respectively. **b**, Spin-polarized tunnelling spectra under ±0.5 T magnetic fields. Here are displayed the average curves obtained from four sites (Fig. S11). **c**, Magnetic asymmetry ($A_{magn}$) calculated from the experimental curves in **b** (upper panel) and the theoretical spin DOS (lower panel). **d**, Tunnelling spectra zoomed-in on the $P_{-1}$ state at B = ±0.15 T and 0 T, with forward/backward magnetic field sweeps in the left/right panel. **e**, d$I$/d$V$ signals of the $P_{-1}$ state

as a function of the magnetic field at 0.35 K (upper panel) and 4.3 K (lower panel). The ferromagnetic hysteresis loops are outlined with rectangular solid lines. Red and blue points/curves represent forward and backward sweeps. Insets in **b** and **e** illustrate the magnetization configurations between the tip and the sample. **f**, DFT calculated spin density plot, where the pink (blue) contours denote the spin-up (spin-down). Yellow and blue circles highlight Mo atoms with the largest and second-largest magnetic moments, respectively. Stabilization parameters: **b**, $V_{bias}$ = 700 mV, $I_t$ = 200 pA, and $V_{mod}$ = 6 mV; **d** and **e**, $V_{bias}$ = 400 mV, $I_t$ = 200 pA, and $V_{mod}$ = 4 mV.

To confirm the itinerant FM properties of monolayer $Mo_{33}Te_{56}$, we performed spin-polarized STM (SP-STM) measurements with an out-of-plane magnetic field (see Fig. 3a). Antiferromagnetic Cr-coated W tips, with stable tip apex magnetization under varying external magnetic field, were utilized (tip calibrations and other technical details are discussed in Fig. S9 and the Methods section)[40,41]. Figure 3b presents the averaged tunnelling conductance recorded on the four representative sites of an ML-$Mo_{33}Te_{56}$ sample under fields $B$ = ±0.5 T. The spectroscopic intensities varied noticeably across the energy range due to the magnetization flip of the sample moment under the reversal of the external magnetic field. The magnetic asymmetry $A_{magn}$, is a widely used quantity to represent such intensity variation, thus the spin-polarization of samples[42]. We extracted $A_{magn}$ as a function of energy by dividing the subtraction of energy-dependent conductance values under positive and negative fields by the sum of them [Methods] as shown in Fig. 3c[42]. This magnetic asymmetry is proportional to the difference between the spin-up and -down DOS $D_{sp}(E) = D_{up}(E) - D_{down}(E)$, which could be derived from the spin-polarized band structures (Fig. S8c). The appearance of $D_{sp}$(E) reproduces major features of the experimental $A_{magn}$ spectrum. Additionally, the pre-calibrate tip results indicate that the tip magnetization is along the +$z$ direction (Fig. S9). This orientation identification correlates the experimental and theoretical (spin-up) majority spins predominately exhibited in the $P_{-1}$ state, whereas the $P_0$ and $P_1$ states are polarized non-negligibly in the minority spin.

With $B$ continuously swept within -0.2 T to 0.2 T in 0.02 T increments, we conducted further spectroscopic measurements focusing on $P_{-1}$ due to its substantial $A_{magn}$ magnitude (Fig. 3c). Data on other peaks are in the Supporting Information (Fig. S10, S11). Selective spectra focusing on $P_{-1}$ under $B$ = 0, ±0.15 T are shown in Fig. 3d. The conductance value is plotted as a function of $B$ in Fig. 3e. The magnetization curve is slightly shifted due to the residual stray filed of the tip[42]. A rectangular hysteresis loop is observable at either 0.35 or 4.3 K. The coercive field is approximately 0.10 T at 0.35 K (upper panel) and 0.06 T at 4.3 K (lower panel). The ferromagnetic hysteresis behaviour was reliably reproduced using multiple spin-polarized tips, as presented in Fig. S12. Moreover, we further reveal orbital diamagnetism coexisting with spin ferromagnetism through observing the magnetization-polarized-Zeeman effect for the $P_1$ state (Fig. S13), akin to those found in 3D kagome magnets[6-8,43].

A spatial plot of the spin density for the FM state (Fig. 3f) indicates that spin-polarized electrons, contributing to a total magnetization of 1.66 $\mu_B$, are distributed throughout the monolayer, mainly on Mo atoms around the MTBs and shared corners. The Mo atoms highlighted by yellow circles have the largest local magnetic moment of 0.20 $\mu_B$, those by blue circles have a reduced moment of 0.13 $\mu_B$, while other Mo sites have even smaller moments (< 0.10 $\mu_B$). These local magnetic moments are an order of magnitude smaller than those in typical classical magnets. Experimental observations align with this finding, showing delocalized magnetic moments, as evidenced by a similar appearance in the $A_{\mathrm{magn}}$ spectrum across the entire primitive cell (Fig. S11). The quantitative differences observed at various sites are consistent with the spin density distribution illustrated in Fig. 3f. This behaviour differentiates flat bands with a destructive interference origin from those with suppressed wavefunction overlap, such as the Mott-like flat band in the 1T-$TaSe_2$ CDW monolayer, which is subject to a local moment picture[44,45] (Fig. S11).

Alongside the itinerant ferromagnetism, another intriguing feature is the puzzling dip at the Fermi level. Figure 4a presents high-resolution tunnelling spectra within a zoomed-in energy range (±100 meV) taken at 0.35 K. A hard bandgap of $\Delta_0 \sim 15$ meV for the zero-conductance region is clearly observed. This bandgap-width magnitude persists across different locations, although minor differences in the line shapes occur. To identify the origin of this hard bandgap, we measured its evolution with temperature. As shown in Fig. 4b, the hard bandgap continuously narrows with increasing temperature and closes above 12 K. However, this bandgap closing cannot be explained solely by thermal broadening. In Fig. 4c, the 0.35 K spectrum (black curve) was used to simulate the thermal broadening at $T = 20$ K (red curve; see Supplementary Note 3 for details). The simulated 20 K data significantly deviates from the measured 20 K spectrum (grey curve). To quantify the gap size over a wide temperature range, we define $\Delta_N$ by the midpoints of the edge slopes. Figure 4d shows the simulated thermal broadening of $\Delta_N$ and the experimentally measured values as a function of temperature, where the experimental results exhibit a more rapid decay. This spectroscopic behaviour indicates the correlated nature of the observed hard bandgap[46,47], distinct from the single-particle gap observed in monolayer $Mo_5Te_8$[22], a cousin monolayer of $Mo_{33}Te_{56}$.

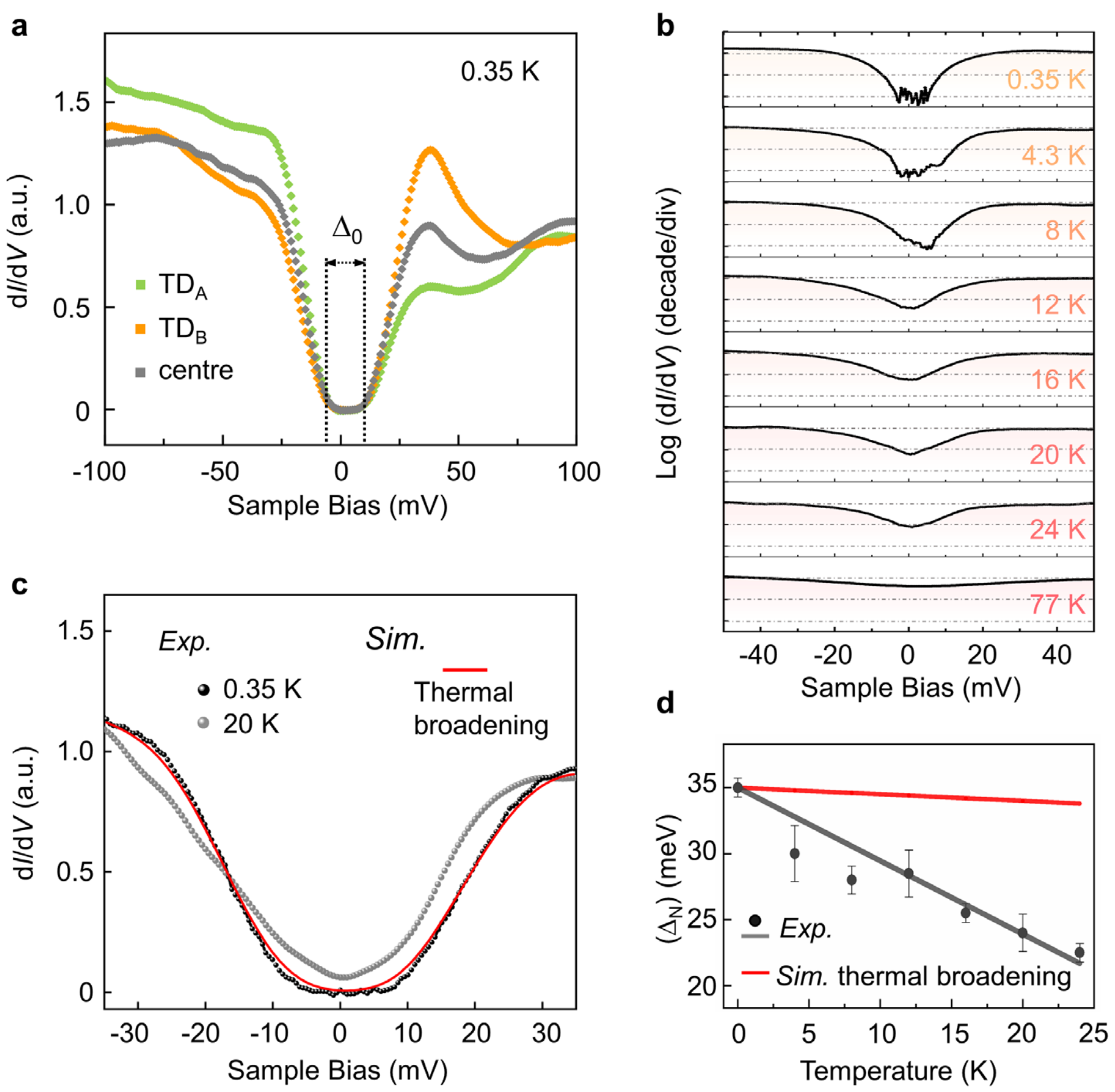

**Fig. 4 | Experimental evidence for the correlated gap at the Fermi level. a**, High-resolution d$I$/d$V$ spectra recorded at the $TD_{E/U}$ and centre sites at $T$ = 0.35 K, showing a relatively uniform hard energy gap. $\Delta_0$ denotes the size of the zero-conductance region (Fig. S16). **b**, Temperature-dependent d$I$/d$V$ spectra (centre site) from 0.35 K – 77 K, plotted on a logarithmic scale. **c**, Selective spectra taken at 0.35 K (black curve) and 20 K (grey curve). The red curve is a thermal broadening simulation ($T$ = 20 K) of the 0.35 K spectrum. **d** shows the experimental $\Delta_N$ (black spots) at different sample temperatures. Grey line is a visual guide, and the red line represents the simulated $\Delta_N(T)$ from thermal broadening of the experimental 0.35 K spectrum (Supplementary Note 3). $\Delta_N$ is the gap defined by the midpoints of the edge slopes, *i.e.*, the maximum point in the $|d^2I/dV^2|$ spectra in **c**. Stabilization parameters: $V_{bias}$ = 100 mV, $I_t$ = 200 pA, and $V_{mod}$ = 2 mV.

In Supplementary Note 4, we elaborate that the observed bandgap is not a Coulomb gap, typically caused by a repulsive action during the electron tunnelling process. This assertion is based on the spectroscopic line shape (*i.e.*, a hard bandgap) and the independence of gap size on the flake area (Fig. S14). A CDW bandgap can also be ruled out, as supplementary experimental results show no charge modulation in the bias-dependent STM imaging (Fig. S14). Furthermore, the possibility of a superconducting gap is excluded, as the gap remains unsuppressed by applying magnetic fields up to 12 T. (Fig. S15). While these common scenarios have been ruled out, pinpointing a precise model to describe the states of matter arising from the partially filled flat bands with specific filling factors and interaction parameters remains challenging[48,49]. This complexity is further exacerbated by the large unit cell and complex orbital composition in the present monolayer, necessitating substantial additional research efforts that integrate multiple experimental techniques[15,17] beyond STM and more advanced theoretical methods than DFT+*U*. Such investigation will be pursued in follow-up work, as it is beyond the scope of the present study that primarily introduces a new single-layer crystalline structure with FB-driven correlated physics.

In conclusion, we have developed a monolayer kagome system with FB-driven correlated magnetism in monolayer $Mo_{33}Te_{56}$, derived from H-phase $MoTe_2$ with a highly controllable and scalable MTB loop superlattice. Our theoretical calculations validate the presence of multiple kagome flat bands within this superlattice. In the partially filled bands, strong electron correlations induce a bandgap opening up to 15 meV. The spin-polarized STS provides strong evidence of spontaneous spin polarization. Future studies are expected to explore the doping dependence of the correlated insulator by tuning the carrier density. This work introduces a novel strategy for designing monolayer vdW materials with pronounced correlation effects, unlocking versatile opportunities for engineering nontrivial emergent quantum phenomena.

## Methods

**Sample preparation.** In this study, samples were grown on bilayer graphene formed by heating SiC (0001) in a home-built MBE system with a base pressure of ~ $1.2 \times 10^{-10}$ Torr. Te (99.999%) was evaporated from Knudsen cells, and High-purity Mo (99.95%) was evaporated from an e-beam evaporator, respectively. The flux ratios of Mo/Te were 1:30. To obtain the ordered MTB superstructure, one follows the following procedure. The substrate was kept at ~ 250 °C during the deposition. After deposition, the sample was annealed for 10 min at the growth temperature with the Te flux maintained. Subsequently, the substrate temperature was increased to 450 °C and annealed for 1 hour (with the Te flux off).

**STM/ Q-plus AFM measurements.** The STM/STS measurements were performed by a commercial Unisoku 1300 system (base pressure < $1\times10^{-10}$ Torr). Electrochemically etched W-tips were used in all measurements, which are calibrated spectroscopically against the Shockley surface states of cleaned Cu (111) surfaces before performing measurements on $MoTe_2$. STM topographic images are acquired in constant-current mode. The d$I$/d$V$ spectra are measured using a standard lock-in amplifier with modulation at a frequency of 932 Hz; other parameters are specified in figure captions. All STM/S measurements were taken at 4.3 K unless otherwise specified.

The nc-AFM imaging is performed by the frequency shift ($\Delta f$) of the qPlus resonator in constant-height mode with an oscillation amplitude of 100 pm. The resonance frequency of the AFM probe is $f_0$ = 29.10 kHz, and its quality factor Q is 57962.

For spin-polarized STM measurements, a W tip was coated with ~20 atomic layers of Cr (purity: 99.995%) after being flashed to 2200 K to remove oxides. The tip was then annealed at ~600 K for 10 minutes to recrystallize the Cr coating, stabilizing the structural phase at the tip apex. This process generated an antiferromagnetic tip with an out-of-plane net magnetic moment and high coercivity, making it stable against external magnetic fields[40,41]. The spin-polarized properties of the Cr-coated W tips were verified using Co islands on Cu (111) following standard calibration procedures (Fig. S9).

The magnetic asymmetry ($A_{magn}$) is calculated using the formula: $A_{magn} = [(dI/dV)_{+B} - (dI/dV)_{-B})] / [(dI/dV)_{+B} + (dI/dV)_{-B}]$[42], where $(dI/dV)_{+B}$ and $(dI/dV)_{-B}$ correspond to the conductance for $+B$ = +0.5 T and $-B$ = -0.5 T, respectively, as shown in Fig. 3b.

**DFT calculation.** The theoretical calculations were performed using the VASP package based on DFT. The electron-ion potential and exchange-correlation functional were described by projected augmented wave (PAW) and generalized gradient 0approximation (GGA)[50,51], respectively. For the relaxation of geometric structures, the kinetic energy cutoff is set to 350 eV. A 3×3×1 Monkhorst Pack k-point grid was used for structure optimization, whereas a denser 7×7×1 grid for electronic calculation[52]. As a double-check, the energy cutoff was set to 500 eV and a 5×5×1 k-mesh was used to verify the calculation results. The vacuum region of 20 Å was used to avoid the periodic interaction. The stress force and energy convergence criteria are chosen as 0.01 eV/Å and $10^{-5}$ eV, respectively. All atoms were allowed to be fully optimized to the ground state without considering van der Waals dispersion corrections. The band decomposed charge density was calculated by reading the converged wave-function file. The irreps calculations were performed using the IRVSP program in conjunction figure with VASP[53].

The DFT simulation of the frequency shift AFM image was obtained using previously established methods, as detailed in ref. 54. A 3D field of the frequency shift[55] was captured with a resolution of 20 × 20 × 1 data points in a box of 19.13 Å × 19.13 Å × 25Å above the $Mo_{33}Te_{56}$ sample.

## Acknowledgements

This work was supported by the National Key R&D Program of China (Grant No. 2018YFA0703700, No. 2018YFE0202700, No. 2022YFA1402401), the National Natural Science Foundation of China (12134011, 12174291, 11974012, 11974422, and 12204534), the fellowship of China Postdoctoral Science Foundation (Grant No. 2021M702532) and the Strategic Priority Research Program of Chinese Academy of Sciences (XDB30000000). the Fundamental Research Funds for the Central Universities, and the Research Funds of Renmin University of China [Grants No. 22XNKJ30 (W.J.) and No. 23XNH077 (J.D.)]. All numerical calculations presented in this paper were performed on the supercomputing system in the Supercomputing Center of Wuhan University.

## References

1. Derzhko, O., Richter, J. & Maksymenko, M. Strongly correlated flat-band systems: the route from Heisenberg spins to Hubbard electrons. *Int. J. Mod. Phys. B* **29**, 1530007 (2015).
2. Regnault, N. et al. Catalogue of flat-band stoichiometric materials. *Nature* **603**,

824–828 (2022).

3. Bergman, D. L., Wu, C. & Balents, L. Band touching from real-space topology in frustrated hopping models. *Phys. Rev. B* **78**, 125104 (2008).
4. Ye, L. et al. Massive Dirac fermions in a ferromagnetic kagome metal. *Nature* **555**, 638–642 (2018).
5. Kang, M. et al. Dirac fermions and flat bands in the ideal kagome metal FeSn. *Nat. Mater.* **19**, 163–169 (2020).
6. Yin, J.-X. et al. Negative flat band magnetism in a spin–orbit-coupled correlated kagome magnet. *Nat. Phys.* **15**, 443–4482 (2019).
7. Xing, Y. et al. Localized spin-orbit polaron in magnetic Weyl semimetal $Co_3Sn_2S_2$. *Nat. Commun.* **11**, 5613 (2020).
8. Huang, H. et al. Flat-band-induced anomalous anisotropic charge transport and orbital magnetism in kagome metal CoSn. *Phys. Rev. Lett.* **128**, 096601 (2022).
9. Li, M. et al. Dirac cone, flat band and saddle point in kagome magnet $YMn_6Sn_6$. *Nat. Commun.* **12**, 31292 (2021).
10. Sun, Z. et al. Observation of topological flat bands in the kagome semiconductor $Nb_3Cl_8$. *Nano Lett.* **22**, 4596–4602 (2022).
11. Wang, L. et al. Correlated electronic phases in twisted bilayer transition metal dichalcogenides. *Nat. Mater.* **19**, 861–866 (2020).
12. Abbas, G. et al. Recent advances in twisted structures of flatland materials and crafting moiré superlattices. *Adv. Funct. Mater.* **30**, 2000878 (2020).
13. Li, T. et al. Continuous Mott transition in semiconductor moiré superlattices. *Nature* **597**, 350–354 (2021).
14. Anderson, E. et al. Programming correlated magnetic states with gate-controlled moiré geometry. *Science* **381**, 325–330 (2023).
15. Cao, Y. et al. Correlated insulator behaviour at half-filling in magic-angle graphene superlattices. *Nature* **556**, 80–84 (2018).
16. Xie, M. & MacDonald, A. H. Nature of the correlated insulator states in twisted bilayer graphene. *Phys. Rev. Lett.* **124**, 097601 (2020).
17. Liu, X. et al. Tunable spin-polarized correlated states in twisted double bilayer

graphene. *Nature* **583**, 221–225 (2020).

18. Tschirhart, C. L. et al. Imaging orbital ferromagnetism in a moiré Chern insulator. *Science* **372**, 1323–1327 (2021).
19. Cao, Y. et al. Unconventional superconductivity in magic-angle graphene superlattices. *Nature* **556**, 43-50 (2018).
20. Oh, M. et al. Evidence for unconventional superconductivity in twisted bilayer graphene. *Nature* **600**, 240–245 (2021).
21. Liu, H., Meng, S. & Liu, F. Screening two-dimensional materials with topological flat bands. *Phys. Rev. Mater.* **5**, 084203 (2021).
22. Lei, L. et al. Electronic Janus lattice and kagome-like bands in coloring-triangular $MoTe_2$ monolayers. *Nat. Commun.* **14**, 6320 (2023).
23. Zhang, H. et al. Topological Flat Bands in 2D Breathing-Kagome Lattice $Nb_3TeCl_7$. *Adv. Mater.* **35**, 2301790 (2023).
24. Liu, H. et al. Dense network of one-dimensional midgap metallic modes in monolayer $MoSe_2$ and their spatial undulations. *Phys. Rev. Lett.* **113**, 066105 (2014).
25. He, X. et al. Selective self-assembly of 2,3-diaminophenazine molecules on $MoSe_2$ mirror twin boundaries. *Nat. Commun.* **10**, 2847 (2019).
26. Zhu, T. et al. Imaging gate-tunable Tomonaga–Luttinger liquids in 1H-$MoSe_2$ mirror twin boundaries. *Nat. Mater.* **21**, 748–753 (2022).
27. Chen, J. et al. Quantum effects and phase tuning in epitaxial hexagonal and monoclinic $MoTe_2$ monolayers. *ACS Nano* **11**, 3282–3288 (2017).
28. Dong, Lu, et al. Charge density wave states in 2H-$MoTe_2$ revealed by scanning tunneling microscopy. *Chin. Phys. Lett.* **35**, 066801 (2018).
29. Yu, Y. et al. Phase-controlled growth of one-dimensional $Mo_6Te_6$ nanowires and two-dimensional $MoTe_2$ ultrathin films heterostructures. *Nano Lett.* **18**, 675–681 (2018).
30. Barja, S. et al. Charge density wave order in 1D mirror twin boundaries of single-layer $MoSe_2$. *Nat. Phys.* **12**, 751–756 (2016).
31. Zhu, H. et al. Defects and surface structural stability of $MoTe_2$ under vacuum

annealing. *ACS Nano* **11**, 11005–11014 (2017).

32. Batzill, M. Mirror twin grain boundaries in molybdenum dichalcogenides. *J. Phys. Condens. Matter* **30**, 493001 (2018).
33. Zhang, J. et al. Single-layer $Mo_5Te_8$—A new polymorph of layered transition-metal chalcogenide. *2D Mater.* **8**, 015006 (2020).
34. Park, J. W., Cho, G. Y., Lee, J. & Yeom, H. W. Emergent honeycomb network of topological excitations in correlated charge density wave. *Nat. Commun.* **10**, 4038 (2019).
35. Kruthoff, J., de Boer, J., van Wezel, J., Kane, C. L. & Slager, R.-J. Topological classification of crystalline insulators through band structure combinatorics. *Phys. Rev. X* **7**, 041069 (2017).
36. Bradlyn, B. et al. Topological quantum chemistry. *Nature* **547**, 298–305 (2017).
37. Elcoro, L. et al. Double crystallographic groups and their representations on the Bilbao Crystallographic Server. *J. Appl. Crystallogr.* **50**, 1457–1477 (2017).
38. Zhuang, H. L., Kent, P. R. C. & Hennig, R. G. Strong anisotropy and magnetostriction in the two-dimensional Stoner ferromagnet $Fe_3GeTe_2$. *Phys. Rev. B* **93**, 134407 (2016).
39. Kang, L. et al. Phase-controllable growth of ultrathin 2D magnetic FeTe crystals. *Nat. Commun.* **11**, 3729 (2020).
40. Chen, W. et al. Direct observation of van der Waals stacking–dependent interlayer magnetism. *Science* **366** 983-987 (2019).
41. Xian, J. et al. Spin mapping of intralayer antiferromagnetism and field-induced spin reorientation in monolayer $CrTe_2$. *Nat. Commun*, **13**, 257 (2022).
42. Meier, F. et al. Revealing magnetic interactions from single-atom magnetization curves. *Science* **320**, 82-86 (2008).
43. Li, H. et al. Spin Berry curvature-enhanced orbital Zeeman effect in a kagome metal. *Nat. Phys.* 1–7 (2024).
44. Ruan, W. et al. Evidence for quantum spin liquid behaviour in single-layer 1T-$TaSe_2$ from scanning tunnelling microscopy. *Nat. Phys.* **17**, 1154–1161 (2021).
45. Liu, M. et al. Monolayer 1T-$NbSe_2$ as a 2D-correlated magnetic insulator. *Sci. Adv.*

**7**, eabi6339 (2021).

46. Vano, V. et al. Artificial heavy fermions in a van der Waals heterostructure. *Nature* **599**, 582–586 (2021).
47. Yang, X. et al. Possible phason-polaron effect on purely one-dimensional charge order of $Mo_6Se_6$ nanowires. *Phys. Rev. X* **10**, 031061 (2020).
48. Tasaki, H. From Nagaoka's ferromagnetism to flat-band ferromagnetism and beyond: an introduction to ferromagnetism in the Hubbard model. *Prog. Theor. Phys.* **99**, 489–548 (1998).
49. Liu, Z., Liu, F. & Wu, Y.-S. Exotic electronic states in the world of flat bands: From theory to material. *Chin. Phys. B* **23**, 077308 (2014).
50. Perdew, J. P., Burke, K. & Ernzerhof, M. Generalized gradient approximation made simple. *Phys. Rev. Lett.* **78**, 1396–1396 (1997).
51. Blöchl, P. E. Projector augmented-wave method. *Phys. Rev. B* **50**, 17953–17979 (1994).
52. Monkhorst, H. J. & Pack, J. D. Special points for Brillouin-zone integrations. *Phys. Rev. B* **13**, 5188–5192 (1976).
53. Gao, J., Wu, Q., Persson, C. & Wang, Z. Irvsp: To obtain irreducible representations of electronic states in the VASP. *Comput. Phys. Commun.* **261**, 107760 (2021).
54. Hapala, P. et al. Mechanism of high-resolution STM/AFM imaging with functionalized tips. *Phys. Rev. B* **90**, 085421. (2014).
55. Gross, L. et al. The chemical structure of a molecule resolved by atomic force microscopy. *Science*, **325**, 1110-1114. (2009).